\documentclass[12pt]{article}
\usepackage{amsmath}
\usepackage{graphics}

\makeatletter

\@addtoreset{equation}{section}
\makeatother

\begin{document}
\baselineskip=16pt
\begin{titlepage}
\begin{flushright}
KUNS-1935, KYUSHU-HET-74\\[-1mm] hep-ph/0409355
\end{flushright}
\begin{center}
\vspace*{1.3cm}

{\Large\bf%
Radiative stabilization of warped space%
}\vspace{9mm}

Tatsuo Kobayashi$^{\rm a}$ and Koichi Yoshioka$^{\rm b}$
\vspace*{3mm}

$^{\rm a}$ Department of Physics, Kyoto University, Kyoto 606-8502, Japan\\
$^{\rm b}$ Department of Physics, Kyushu University, Fukuoka 812-8581, Japan
\vspace*{2mm}

{\small (September, 2004)}
\end{center}
\vspace*{5mm}

\begin{abstract}\noindent%
Higher-dimensional field theory has been applied to explore various
issues in recent particle physics such as the gauge hierarchy
problem. In order for such approaches to be viable, a crucial
ingredient is to fix the sizes of extra dimensions at some finite
values, which sizes are generically free parameters in the theory. In
this paper, we present several schemes to determine the radius of
extra dimension in warped five-dimensional theory. In every case, a
non-vanishing Fayet-Iliopoulos term for abelian gauge factor plays a
crucial role for the radius stabilization. It is radiatively generated
in the presence of charged matter fields and the compactification is
therefore spontaneous, not forced by selected operators. The
low-energy supersymmetry is broken or unbroken, and the radius can be
fixed to give a small or large scale hierarchy without any fine tuning
of parameters. We also discuss a model of the radius stabilization
correlated with Yukawa hierarchy and supersymmetry breaking.
\end{abstract}
\end{titlepage}

\section{Introduction}

Field theory in higher dimensions has been providing novel approaches
to theoretical and phenomenological problems in recent particle
physics. The existence of extra spatial dimensions beyond our fours is
applied to various issues such as the generation of large scale
hierarchies~\cite{ADD,RS}. In these approaches, a key
ingredient is to determine the size of extra space so that they are
viable approaches and do not conflict with current observations. For
example, the Planck/weak mass hierarchy is attained by assuming that
the radii of compactified dimensions are huge~\cite{ADD} or small but
a bit larger than the Planck length~\cite{RS}. The compactification
radius is also conjectured to have anticipated values in other
phenomenological discussions such as small neutrino
masses~\cite{EDneu}, Yukawa hierarchies of quarks and
leptons~\cite{EDYukawa}, and supersymmetry
breaking~\cite{EDsusybreaking}. Therefore adjusting the sizes of extra
dimensions to desired values is one of the most important issues in
higher-dimensional framework. There have been in the literature
various resolutions to this stabilization problem in large- and
small-sized extra dimensions~\cite{models}.

In this paper, we present three different schemes to stabilize the
radius modulus in five-dimensional supersymmetric theory with or
without charged matter fields. One is based on the model with only
boundary charged fields and another with only bulk fields. In every
scheme, the Fayet-Iliopoulos (FI) term~\cite{FID} in five-dimensional
theory is found to play a crucial role for the radius
stabilization. The resultant metric factor can be significant or
nearly flat, depending on model parameters. It is stressed that the FI
term is not introduced by hand in order to stabilize the radius. A
non-vanishing FI term is radiatively generated even if it is set to
zero in the classical Lagrangian~\cite{FIexD}. This is unlike the
four-dimensional theory where a FI term does not receive any
renormalization if theory has a vanishing gravitational
anomaly~\cite{NRT-D}. The induced FI term depends on how charged
matter multiplets are distributed in the extra dimensions and is
therefore controllable. Further it is known in four-dimensional models
that the FI term is connected to Yukawa hierarchy and supersymmetry
breaking~\cite{anomalousU1}. We construct a five-dimensional model for
generating fermion mass hierarchy which is correlated with the radius
stabilization. The model also predicts characteristic spectrum of
sfermions and gauginos.

This paper is organized as follows. In Section 2, we give a generic
form of globally supersymmetric Lagrangian for 
five-dimensional $U(1)$ theory with the FI term. The simplest
stabilization scheme is found in Section 3 without introducing any
matter fields. In Section 4, a different scheme is presented to fix
the size of extra dimension. The model contains bulk hypermultiplets
with non-trivial wavefunctions whose forms are determined by FI-term
coefficients. The bulk multiplets fix the distance between the two
boundaries in terms of boundary couplings. In the vacuum of this
model, supersymmetry is unbroken. With the generic Lagrangian at hand,
we show in Section 5 that the radius can be stabilized by only boundary
matter fields. That is established by analyzing the vacuum energy in
effective four-dimensional theory with broken supersymmetry. We also
construct a toy model for Yukawa hierarchy and supersymmetry breaking,
deeply correlated with the radius stabilization phenomenon. Such a
model predicts a new type of sparticle spectrum in low-energy
effective theory. Section 6 is devoted to summarizing our results.

\section{Five-dimensional $U(1)$ gauge theory}

We consider the globally supersymmetric abelian gauge theory in five
dimensions. The fifth dimension is compactified on a line segment
$S^1/Z_2$, where the radius of the circle $S^1$ is $R$. The radius $R$
is a free parameter of the theory and corresponds to a massless moduli
field $T$ in four-dimensional effective theory
($R\equiv\textrm{Re}\,T$). The fifth dimension $y$ has two boundaries
at $y=0$ and $\pi R$. They are fixed points under the $Z_2$
orbifolding of physical spacetime. We are now interested in the case
that the extra dimension has curved geometry. A particularly
interesting example is the warped (AdS) geometry~\cite{RS} whose line
element is given by
\begin{equation}
  ds^2 \,=\, e^{-2k|y|}\eta_{\mu\nu}dx^{\mu}dx^{\nu}+dy^2,
  \label{warp}
\end{equation}
where $k$ is the AdS curvature and $\eta_{\mu\nu}$ the Minkowski
metric in four dimensions. This background metric has been intensively
studied for realistic model construction with the Planck/weak scale
difference, quarks and leptons mass hierarchy, the cosmological
constant problem, etc. In these approaches the radius $R$ of the
compact extra dimension was often assumed to have a desired value, and
the radius stabilization is therefore one of the most important
problems in constructing realistic `brane-world' models. In this paper
we present the schemes to stabilize $R$ at a finite value due to the
existence of $U(1)$ gauge factor in supersymmetric warped dimensions.

We adopt the superspace formalism of higher-dimensional
supersymmetry~\cite{sf}. There are two types of supermultiplets
generally introduced in five-dimensional theory; vector and hyper
multiplets. A vector multiplet contains an $N=1$ vector multiplet $V$
and a chiral multiplet $\chi$, whose auxiliary components are denoted
by $D$ and $F_\chi$, respectively. A hypermultiplet consists of
oppositely-charged two chiral multiplets $\phi$ and $\phi^c$. In the
superspace language, the most generic Lagrangian for five-dimensional
$U(1)$ gauge theory is given by
\begin{equation}
  L \;=\; L_V+L_H+L_{\rm UV}\delta(y)+L_{\rm IR}\delta(y-\pi R)+L_D,
\end{equation}
\begin{eqnarray}
  L_V &\!=\!& \int\! d^2\theta\,\frac{1}{4g^2}
  W^\alpha W_\alpha +{\rm h.c.} +\int\! d^4\theta\,
  \frac{e^{-2k|y|}}{g^2}\Big[\partial_yV-\frac{1}{\sqrt{2}}
  (\chi+\chi^\dagger)\Big]^2, \label{LV} \\[1mm]
  L_H &\!=\!& \int\! d^4\theta\, e^{-2k|y|} \big(\phi^\dagger
  e^{qV}\phi +\phi^c e^{-qV} {\phi^c}^\dagger \big) \nonumber \\
  && \hspace{2cm} +\int\! d^2\theta\, e^{-3k|y|} 
  \phi^c\bigg[\partial_y+\frac{q}{\sqrt{2}}\chi 
  -\Big(\frac{3}{2}-c\Big)k\epsilon(y)\bigg]\phi +{\rm h.c.}, \label{LH}
\end{eqnarray}
where $g$ denotes the gauge coupling constant, and $q$ and $c$ are 
the $U(1)$ charge and the bulk mass parameter of the chiral 
multiplet $\phi$, respectively. The sign function $\epsilon(y)$ is
inserted in order for the orbifold $Z_2$ invariance. We have also
included the Lagrangian for chiral multiplets confined on the UV
($y=0$) and IR ($y=\pi R$) boundaries;
\begin{eqnarray}
  L_{\rm UV} &\!=\!& \int\! d^4\theta\, \phi_{\rm UV}^\dagger 
  e^{q_{\rm UV}V}\phi_{\rm UV} +\int\! d^2\theta\, 
  W_{\rm UV}(\phi,\phi_{\rm UV}) +{\rm h.c.}, \label{LUV} \\
  L_{\rm IR} &\!=\!& \int\! d^4\theta\, e^{-2k\pi R}
  \phi_{\rm IR}^\dagger e^{q_{\rm IR}V}\phi_{\rm IR} 
  +\int\! d^2\theta\, e^{-3k\pi R} W_{\rm IR}(\phi,\phi_{\rm IR})
  +{\rm h.c.}. \label{LIR}
\end{eqnarray}
The orbifold boundary conditions are imposed on each
supermultiplet. The vector multiplet $V$ has the Neumann boundary
conditions at both UV and IR branes and its superpartner 
multiplet $\chi$ has the Dirichlet ones because it contains the fifth
component of the bulk gauge field. The boundary conditions 
of $\phi$ must be opposite to those of superpartner $\phi^c$ for
respecting the $Z_2$ symmetry. The $Z_2$ boundary conditions break a
half of bulk supersymmetry and thus the boundary 
Lagrangians $L_{\rm UV}$ and $L_{\rm IR}$ preserve only the
four-dimensional $N=1$ supersymmetry. For example, Yukawa couplings of 
quarks and leptons are expected to come from these boundary
interactions in the present framework. The boundary chiral 
multiplets $\phi_{\rm UV}$ and $\phi_{\rm IR}$ couple only to bulk
multiplets with Neumann boundary conditions. We have assumed, for
simplicity, that there are no $y$-derivative couplings of $Z_2$-odd
chiral multiplets and no four-dimensional gauge fields on the
boundaries, while these assumptions are irrelevant to the following
discussion. The exponential warp factors are explicitly included in
the above Lagrangian. These warp factors describe the metric
dependences, such as from $\sqrt{-\det g_{\mu\nu}}$, of the lowest
component of each supermultiplet in the warped background. For other
component fields, the proper metric factors in the warped
five-dimensional action are obtained after some rescaling, for 
example, $D\to e^{-2k|y|}D$, $F_\phi\to e^{-k|y|}F_\phi$ for the
auxiliary fields.

Since we now consider the abelian gauge theory, a FI term of vector
multiplet $V$ is gauge invariant in globally supersymmetric theory and
can also be added to the Lagrangian as
\begin{equation}
  L_D \,=\, \int\! d^4\theta\, 2\xi V.
  \label{LD}
\end{equation}
We have defined the coefficient $\xi$ into which the metric warp
factor is absorbed. Even if there is no FI term in classical
Lagrangian, it is radiatively generated via tadpole graphs of the $D$
component where charged matter fields circulate in the loop. In the
case of flat extra dimensions, its form was
investigated~\cite{FIexD,FIexD2,FIexD3} and found to reside only on
the orbifold fixed points. Moreover the FI term vanishes in
anomaly-free low-energy effective theory when one integrates out the
fifth-dimensional physics. This is consistent with the fact that, in
four-dimensional theory, a coefficient of radiatively-generated FI
term is proportional to the sum of matter $U(1)$ charges which also
gives the mixed $U(1)$-gravitational anomaly. In five-dimensional
theory on curved backgrounds including the warped geometry, the
situation is rather different. Since the fifth direction is curved,
the fundamental length depends on the position $y$. The implication of
this fact appears through the metric-factor dependences in the FI-term
calculation. In the warped geometry~(\ref{warp}), the brane-localized
FI terms are written as
\begin{equation}
  \xi \,=\, \xi_{\rm UV}\delta(y)-\xi_{\rm IR}e^{-2k\pi R}\delta(y-\pi R),
  \label{FI}
\end{equation}
with the constant coefficients $\xi_{\rm UV}$ and $\xi_{\rm IR}$. If
the FI term is set to vanish at classical level, radiative
corrections give rise to $\xi_{\rm UV}$ and $\xi_{\rm IR}$ which are
given by specific combinations of $U(1)$ charges of bulk and boundary
fields~\cite{FIexDwarp}. For a $U(1)$ factor free from gravitational
anomaly in low-energy theory, the two coefficients are equal to each
other; $\xi_{\rm UV}=\xi_{\rm IR}$. The exponential factor in the
second term of (\ref{FI}) indicates that the fundamental length is
redshifted at the $y=\pi R$ boundary. This factor may be described by
proper regularization, for example, \`a la Pauli-Villars, and more
simply implemented by a position-dependent cutoff for the
four-dimensional momentum in the one-loop calculations, which
dependence is suggested by the AdS/CFT 
correspondence~\cite{ADS-CFT}.\footnote{A position-dependent value of
FI term can also be seen from the theory-space approach to
five-dimensional curved backgrounds~\cite{ADS4D}.} That has been
recently confirmed by detailed analysis of five-dimensional
supergravity~\cite{FIDsugra}. In four-dimensional effective theory,
the FI term does not vanish as the zero modes of vector multiplet have
flat wavefunctions. As a result, either $U(1)$ gauge symmetry or
four-dimensional supersymmetry is broken at the scale of $\xi$. This
reflects the known fact in four-dimensional theory that a FI term for
anomaly-free $U(1)$ gauge theory does not coexist with unbroken
supersymmetry. That is, in four-dimensional supergravity theory, a FI
term can be introduced only when $U(1)$ is $R$ symmetry or
non-linearly realized, i.e.\ the $U(1)$ gauge boson becomes
massive. An important point here is that even if low-energy theory is
totally free of anomalies like QED, the effective FI term is
non-vanishing due to the curved extra dimension ($k\neq0$) and has
important phenomenological implications~\cite{FIexDwarp}. In this
paper, we show that the presence of FI term also provides the
stabilization mechanisms of the radius modulus field.

\section{Stabilization without bulk/boundary fields}

Let us first see the simplest case where we have no charged matter
fields in the theory. Integrating out the fifth-dimensional physics,
we find that the presence of the FI term (\ref{FI}) leads to the
potential
\begin{equation}
  V(R) \,=\, \frac{g^2}{4\pi R}
  \big(\xi_{\rm UV}-\xi_{\rm IR}e^{-2k\pi R}\big)^2.
  \label{VR-UVIR}
\end{equation}
This vacuum energy depends on the radius $R$. It is therefore
determined so that the vacuum energy is minimized. We find from
(\ref{VR-UVIR}) a possibility that the radius is fixed to a finite value
\begin{equation}
  kR \,=\, \frac{1}{2\pi}\ln\Big(\frac{\xi_{\rm IR}}{\xi_{\rm UV}}\Big).
  \label{kR}
\end{equation}
The vacuum energy vanishes at this point which is the potential
minimum in globally supersymmetric theory. Thus four-dimensional
supersymmetry is unbroken while the radius is stabilized. Note that
the limit $R\to\infty$ also gives a vanishing vacuum energy, where the
low-energy gauge theory becomes a free theory. However the potential
barrier between the two minima can be as high as $\xi_{\rm UV}^2$
whose natural size is around the Planck scale. Therefore the vacuum
(\ref{kR}) might be made stable within the present age of the
universe. Moreover the parameter region far away from the origin could
be lifted by supersymmetry breaking which we have not included
here. It is noticed that the existence of the vacuum (\ref{kR})
calls a restriction on the FI-term 
coefficients; $\xi_{\rm IR}/\xi_{\rm UV}>1$. For example, in case that
low-energy theory is anomaly free, we 
have $\xi_{\rm UV}=\xi_{\rm IR}$ and hence the radius is not settled
at a finite value. For radiatively-generated FI terms, the 
inequality $\xi_{\rm IR}\neq\xi_{\rm UV}$ is realized 
with `anomalous' matter content. Then one should assume some anomaly
cancellation mechanism that does not affect the FI terms. In this
paper, we do not pursue such a possibility further. Instead we will
discuss the radius stabilization with bulk/boundary matter fields in
the presence of `non-anomalous' FI 
term: $\xi_{\rm UV}=\xi_{\rm IR}\equiv\xi_{\rm FI}$.

\section{Stabilization with bulk fields}

In this section we present a scheme for stabilizing the size of warped
extra dimension which involves only bulk hypermultiplets. A
hypermultiplet which has non-trivial profile of bulk wavefunction
connects two localized FI terms, and determines the distance between
the boundaries. In this way the radius is fixed by the equations of
motion of bulk fields together with their boundary conditions on the
branes.

We consider the five-dimensional $U(1)$ gauge theory with
non-vanishing boundary FI terms. Its Lagrangian is given by (\ref{LV})
and (\ref{LH}) as well as boundary superpotentials $W_{\rm UV}$ 
and $W_{\rm IR}$ for bulk chiral multiplets with even $Z_2$ parity. We do
not include any boundary supermultiplets on the branes. The
five-dimensional scalar potential is generally given by
\begin{equation}
  V_{5\rm D} \,=\, \frac{1}{2g^2}D^2 +\frac{e^{-2k\pi R}}{g^2}|F_\chi|^2
  +\sum_{\phi}e^{-2k\pi R}(|F_\phi|^2 +|F_{\phi^c}|^2).
\end{equation}
The auxiliary fields $D$ and $F$'s are expressed in terms of bulk
scalars through their equations of motion (see below). Here we have
simply assumed that the fifth component of the $U(1)$ vector field
does not have a nonzero expectation value.

\subsection{Model}

The model we present in this section contains two bulk hypermultiplets
with the following $U(1)$ charges and bulk masses:
\begin{eqnarray}
  (\phi,\,\phi^c) &: \textrm{$U(1)$ charge of } \phi=+q, 
  \qquad \textrm{bulk mass}=c_\phi, \nonumber \\
  (\varphi,\,\varphi^c) &: \textrm{$U(1)$ charge of } \varphi=-q,
  \qquad \textrm{bulk mass}=c_\varphi. \nonumber
\end{eqnarray}
Notice that two hypermultiplets have opposite $U(1)$ charges so that
they can form a mixing mass term. This is however just a simplifying
assumption. We will mention other choices of $U(1)$ charges in the end
of this section and show that the charge assignment of hypermultiplets
is irrelevant to the radius stabilization mechanism. The boundary
conditions on the branes, namely, the orbifold parities are taken to be
positive for $\phi$ and $\varphi$ (therefore, negative 
for $\phi^c$ and $\varphi^c$), which lead to the zero modes 
of $\phi$ and $\varphi$ in low-energy effective theory.

\subsection{Unperturbed vacuum}

There exists the supersymmetric vacuum in the presence of FI
term~(\ref{FI}) when the bulk scalars take appropriate expectation
values so that the flatness conditions are satisfied. As mentioned in
the previous section, we consider a conceivable case 
that $\xi_{\rm UV}=\xi_{\rm IR}\equiv\xi_{\rm FI}$. Without loss of
generality, we take $q\xi_{\rm FI}>0$ and then find that among the
bulk matter scalars only $Z_2$-even $\varphi$ develops a vacuum
expectation value. The $D$ and $F$ flatness conditions now reduce to
\begin{eqnarray}
  0 &=& -\partial_y(e^{-2k|y|}\Sigma) 
  +\frac{q g^2e^{-2k|y|}}{2}|\varphi|^2 -g^2\xi_{\rm FI} 
  \big[\delta(y)-e^{-2k\pi R}\delta(y-\pi R)\big], \label{Dflat} \\
  0 &=& \Big[\partial_y-\frac{q}{2}\Sigma 
  -\Big(\frac{3}{2}-c_\varphi\Big)k\epsilon(y)\Big]\varphi.
\end{eqnarray}
The field $\Sigma$ is the real part of the neutral scalar in the
chiral multiplet $\chi$. It seems difficult to write down the generic
solutions of these vacuum equations but we can analytically solve them
for a specific value $c_\varphi=\frac{-1}{2}$. In this case, the
solutions $\Sigma_0$ and $\varphi_0$ are given by
\begin{eqnarray}
  \Sigma_0 &=& \frac{4ka_1}{q}\epsilon(y)e^{2k|y|}
  \tan\big(a_1e^{2k|y|}+a_2\big), \label{sol0S} \\
  \varphi_0 &=& \frac{4ka_1}{qg} e^{2k|y|}
  \frac{1}{\cos\big(a_1e^{2k|y|}+a_2\big)}, \label{sol0p}
\end{eqnarray}
where $a_1$ and $a_2$ are the integration constants. These constants
are determined by the boundary FI terms through the $D$-term
equation~(\ref{Dflat}) as
\begin{equation}
  -\frac{8ka_1}{qg^2}\tan\big(a_1+a_2\big) \,=\, \xi_{\rm FI}, \qquad
  -\frac{8ka_1}{qg^2}\tan\big(a_1e^{2k\pi R}+a_2\big)
  \,=\, \xi_{\rm FI} e^{-2k\pi R}. \label{afix}
\end{equation}
The existence of non-vanishing FI terms therefore fixes the unique
supersymmetric vacuum away from the origin of the field space. When the
warp factor is significant ($kR\gg 1$) and a FI term is small
($\xi_{\rm FI}\ll k^2$), the explicit forms of the integration
constants are approximately given by
\begin{equation}
  a_1 \,\simeq\, \Big(\frac{\pi}{2}-\frac{qg^2\xi_{\rm FI}}{4k\pi}\Big)
  \,e^{-2k\pi R}, \qquad a_2 \,\simeq\, \frac{\pi}{2} 
  +\frac{4k\pi}{qg^2\xi_{\rm FI}}\,e^{-2k\pi R}.
\end{equation}
The wavefunctions $\Sigma_0$ and $\varphi_0$ do not have singularities
between the two boundaries. In the following analysis, we adopt this
value $c_\varphi=\frac{-1}{2}$ as an example. It is however stressed
that one may expect similar effects of radius stabilization for other
generic values of the bulk mass parameters. As we will show below, the
only required is the existence of non-trivial unperturbed solutions.

\subsection{Radius determination}

We introduce the following gauge-invariant superpotential terms onto
the boundaries;
\begin{equation}
  W_{\rm UV} \,=\, m_0\phi\varphi, \qquad W_{\rm IR} \,=\, m_\pi\phi\varphi.
  \label{Wmass}
\end{equation}
It is noted that a supersymmetric mass term between $\phi$ 
and $\varphi$ is forbidden by bulk AdS$_5$ supersymmetry and is
forced to be confined on the boundaries, at which bulk supersymmetry
is broken to the four-dimensional one. On the other hand, the mass
terms $\phi\varphi^c$ and $\phi^c\varphi$ are allowed to exist by the
bulk supersymmetry, but in the present model, the $U(1)$ gauge
invariance makes it vanish. Therefore (\ref{Wmass}) is the most
generic superpotential for the matter fields involved. As we will see,
the above potential terms play important roles of acting as the
sources of $\phi$ which, in turn, stabilizes the radius and of lifting
the $\varphi$ direction from the equation of motion of $\phi$.

Let us first see whether supersymmetry is broken in this model by
examining the four-dimensional scalar potential. The $D$-term
contribution to the scalar potential is written in the usual form with
a non-vanishing FI term. The $F$-term contribution comes from the
superpotential obtained by reducing (\ref{Wmass}) to the
four-dimensional zero-mode part. 
Expanding $\phi(x,y)=\phi_4(x)\phi_y(y)$ 
and $\varphi(x,y)=\varphi_4(x)\varphi_y(y)$, it is given by
\begin{equation}
  \quad\qquad W_{4\rm D} \,=\, \big[ m_0\phi_y(0)\varphi_y(0)-
  m_\pi\phi_y(\pi R)\varphi_y(\pi R)\big] \phi_4(x)\varphi_4(x)
  \>\equiv\, m_{\phi\varphi}\,\phi_4(x)\varphi_4(x),
\end{equation}
where $\phi_y$ and $\varphi_y$ are the solutions of their equations of
motion. If the effective mass parameter $m_{\phi\varphi}$ is 
nonzero, $\phi_4$ and $\varphi_4$ are lifted and supersymmetry is
broken because the $D$-term equation enforces specific values on these
fields. However $m_{\phi\varphi}$ now depends on the radius $R$, which
is generically not a frozen parameter. The radius $R$ thus fixes
itself so as to give a vanishing effective mass $m_{\phi\varphi}$, for
which the vacuum energy is minimized and supersymmetry is restored. In
other words, if the radius modulus $T$ is included as a dynamical
variable, the minimization of scalar potential with respect 
to $\phi_4$, $\varphi_4$ and $T$ leads to a vanishing effective 
mass $m_{\phi\varphi}$ (at least local, supersymmetric vacuum).

Since supersymmetry is unbroken, all the equations of motion in the
five-dimensional theory are
\begin{eqnarray}
  0 \!&=&\! D \,=\, -\partial_y(e^{-2k|y|}\Sigma)
  -\frac{q g^2}{2}e^{-2k|y|}\big(|\phi|^2 -|\phi^c|^2\big) 
  +\frac{q g^2}{2}e^{-2k|y|}\big(|\varphi|^2 -|\varphi^c|^2\big)
  \nonumber\\[1mm]
  &&\hspace*{7cm} -g^2\xi_{\rm FI}\big[\delta(y)
  -e^{-2k\pi R}\delta(y-\pi R)\big],  \label{D} \\
  0 \!&=&\! F_\chi^\dagger \,=\, -\frac{q g^2}{\sqrt{2}}e^{-k|y|}
  (\phi^c\phi-\varphi^c\varphi), \label{Fchi} \\
  0 \!&=&\! F_\phi^\dagger \,=\, e^{-k|y|}\Big[\partial_y
  -\frac{q}{2}\Sigma -\Big(\frac{3}{2}+c_\phi\Big)k\epsilon(y)\Big] \phi^c 
  -\Big[m_0\delta(y)-e^{-k\pi R}m_\pi\delta(y-\pi R)\Big]\varphi, 
  \quad \label{Fphi} \\
  0 \!&=&\! F_\varphi^\dagger \,=\, e^{-k|y|}\Big[\partial_y
  +\frac{q}{2}\Sigma -k\epsilon(y)\Big] \varphi^c 
  -\Big[m_0\delta(y)-e^{-k\pi R}m_\pi\delta(y-\pi R)\Big]\phi, 
  \label{Fvarphi} \\
  0 \!&=&\! F_{\phi^c}^\dagger \!\,=\, -e^{-k|y|} \Big[\partial_y
  +\frac{q}{2}\Sigma -\Big(\frac{3}{2}-c_\phi\Big)
  k\epsilon(y)\Big]\phi, \\
  0 \!&=&\! F_{\varphi^c}^\dagger \!\,=\, -e^{-k|y|} \Big[\partial_y
  -\frac{q}{2}\Sigma -2k\epsilon(y)\Big]\varphi. \label{Fvarphic}
\end{eqnarray}
The localized operators enforce the specific boundary conditions on
the parity-odd functions $\Sigma$, $\phi^c$ and $\varphi^c$ such that
\begin{equation}
  \Sigma \,=\, \epsilon(y) f_\sigma(y), \qquad 
  \phi^c \,=\, \epsilon(y) f_\phi(y), \qquad 
  \varphi^c \,=\, \epsilon(y) f_\varphi(y),
\end{equation}
with the even functions $f_\sigma(y)$, $f_\phi(y)$ and $f_\varphi(y)$
which satisfy the conditions
\begin{eqnarray}
  2f_\sigma(0) &=& -g^2\xi_{\rm FI}, \qquad
  2f_\sigma(\pi R) \;=\; -g^2\xi_{\rm FI}, \\
  2f_\phi(0) &=& m_0\varphi(0), \qquad
  2f_\phi(\pi R) \;=\; m_\pi\varphi(\pi R), \label{bcfp} \\
  2f_\varphi(0) &=& m_0\phi(0), \qquad
  2f_\varphi(\pi R) \;=\; m_\pi\phi(\pi R). \label{bcfvp}
\end{eqnarray}
The $F$-term equations (\ref{Fphi}) and (\ref{Fvarphi}) are simplified
in the bulk as
\begin{eqnarray}
  0 \!&=&\! \Big[\partial_y -\frac{q}{2}\Sigma -\Big(\frac{3}{2}
  +c_\phi\Big) k\epsilon(y)\Big]f_\phi,  \label{fp} \\
  0 \!&=&\! \Big[\partial_y +\frac{q}{2}\Sigma -k\epsilon(y)\Big] f_\varphi.
\end{eqnarray}
Since we have $q\xi_{\rm FI}>0$ without loss of generality, the 
scalar field $\phi$ do not develop vacuum expectation values. This is
also understood from the view of four-dimensional effective theory. In
turn, the equations (\ref{Fchi}) and (\ref{Fvarphi}) together 
with (\ref{bcfvp}) mean $\varphi^c=0$. The independent equations of 
motion now reduce to (\ref{D}), (\ref{Fvarphic}), and (\ref{fp}) 
with $\phi=\varphi^c=0$. 

The vacuum solutions are explicitly derived by solving these equations
in perturbation of $m_0,m_\pi\ll 1$. The leading-order solutions are
given by the unperturbed ones (\ref{sol0S}), (\ref{sol0p}), 
and $f_\phi=0$. It is interesting to notice that the radius
determination does not require a precise form of $\Sigma$, which
follows from the $D$-term equation (\ref{D}). The formal solutions to
(\ref{Fvarphic}) and (\ref{fp}) are
\begin{equation}
  \varphi \,=\, A_\varphi\,e^{2k|y|}
  \exp\big(\mbox{\large $\int$}^y\frac{q}{2}\Sigma\big), \qquad
  f_\phi \,=\, A_\phi\,e^{(\frac{3}{2}+c_\phi)k|y|}
  \exp\big(\mbox{\large $\int$}^y\frac{q}{2}\Sigma\big),
\end{equation}
with $A_\varphi$ and $A_\phi$ being the integration
constants. Inserting the solutions into (\ref{bcfp}), we find that the
boundary conditions of $\phi^c$ determine the value of $R$:
\begin{equation}
  kR \,=\, \frac{\ln\big(\frac{m_\pi}{m_0}\big)}{(c_\phi-\frac{1}{2})\pi}.
  \label{Rstab}
\end{equation}
The boundary conditions also constrain the ratio of integration
constants as $A_\phi/A_\varphi=O(m_0)\ll1$. Their individual values
are fixed by the $D$-term equation which is satisfied 
by $O(m_{0,\pi}^2)$ fluctuation of $\Sigma$ around the unperturbed
solution (\ref{sol0S}). The fact that $A_\phi\ll 1$ ensures the
relevance of the perturbative analysis. The above derivation makes it
clear that explicit solutions to the equations of motion are not
needed to find a stabilized value of the radius. In fact, the radius
in the minimum can easily be evaluated for a generic value 
of $c_\varphi$. That is, a similar analysis shows that $R$ is
determined so that $kR=\ln(\frac{m_\pi}{m_0})/(c_\phi+c_\varphi)\pi$.

Several comments are in order. At least at this order of perturbation,
the stabilized value of the radius does not seem to depend on the FI
term. It is however noticed that the wavefunction factors $A$'s depend
on the FI term via $a_1$. If one turns off the FI 
term ($\xi_{\rm FI}\to 0$), the vacuum goes to the origin of field
space, that is, $a_1\to 0$ [see (\ref{afix})]. Consequently, the
expectation value of  $\phi^c$ also vanishes which cannot lead to the
radius determination (\ref{Rstab}). In this way the existence of
non-vanishing FI term is crucial for the stabilization of radius
modulus field. Secondly, as for the parameters $m_0$ and $m_\pi$,
realizing a significant warp factor does not need any fine tuning. For
example, the values $c_\phi\simeq 0.6$ 
and $\frac{m_\pi}{m_0}\simeq 20$ 
give $e^{k\pi R}\sim 10^{15}$ in (\ref{Rstab}). What is needed is a
parameter choice of order $O(0.1)$, which is similar to that in the
original Randall-Sundrum model where $kR\sim O(10)$ is assumed for
solving the gauge hierarchy problem. Of course, a radius stabilization
with no significant warp factor $e^{k\pi R}\sim O(1)$ is easier to be
achieved. Finally, we comment on the possibility for other choices 
of $U(1)$ charges. When the bulk mass $c_\phi$ is exactly one half,
the equation (\ref{Rstab}) implies that the radius is not
stabilized. This is however simply because of our $U(1)$ charge
assignment. If one supposes the $U(1)$ charge of $\phi$ is $+nq$ for
example, the gauge-invariant boundary superpotentials take the 
form $W\sim\phi\varphi^n$. In this case (also with a general 
unfixed $c_\varphi$), the stabilized value of $R$ is replaced with 
\begin{equation}
  kR \,=\, \frac{\ln(\frac{m_\pi}{m_0})}{(c_\phi+c_\varphi+2-2n)\pi}.
\end{equation}
Thus the radius is still stabilized as long as there is no principle
relating bulk mass parameters and $U(1)$ charges in a specific way.

\section{Stabilization with boundary fields}

In this section we examine whether the radius modulus can be
stabilized only with boundary field dynamics unlike the model
presented in the previous section. In this case, the stabilization
procedure is to look for the minimum of four-dimensional effective
potential of the radius modulus field. Let us first derive low-energy
effective theory for generic boundary superpotential terms. We assume
that supergravity effects except for the radius modulus are irrelevant
to stabilization. Integrating out the fifth dimension, we obtain the
low-energy effective theory of the $U(1)$ multiplet zero modes,
boundary matter multiplets, and the radius modulus $T$. The effective
Lagrangian is derived from (\ref{LV}), (\ref{LUV}), (\ref{LIR}), and
(\ref{LD}) with (\ref{FI}):
\begin{eqnarray}
  {\cal L}_{4\rm D} &\!=\!&\! \int\!d^2\theta\,\frac{\pi T}{2g^2}\, 
  W^\alpha W_\alpha +{\rm h.c.} +\int\!d^4\theta\,
  \Big[ \phi_{\rm UV}^\dagger e^{q_{\rm UV}V}
  \phi_{\rm UV} + e^{-k\pi(T+T^\dagger)}
  \phi_{\rm IR}^\dagger e^{q_{\rm IR}V}\phi_{\rm IR}\Big] \nonumber \\
  && \hspace*{-1mm}
  + \int\!d^2\theta\,\Big[ W_{\textrm{UV}}(\phi_{\rm UV})
  +e^{-3k\pi T} W_{\rm IR}(\phi_{\rm IR}) \Big] +{\rm h.c.} 
  +\!\int\!d^4\theta\Big(2\xi_{\rm FI}V-\frac{6M^3}{k}\Big)
  \big[1-e^{-k\pi(T+T^\dagger)}\big], \nonumber\\[-1mm]
  \label{L4D}
\end{eqnarray}
where $M$ is  the fundamental scale in five-dimensional theory. We
have included the proper K\"ahler term of the radius modulus field in
the warped background~\cite{radionDterm}. It is assumed that
extra bulk dynamics to lead to potential terms of $T$ is not
introduced. As mentioned in the previous section, however, the FI term
is automatically generated in the presence of charged matter fields
and provides $T$-dependent terms. The scalar potential is obtained by
integrating out all the auxiliary components
\begin{eqnarray}
  V_{4\rm D} &=& \frac{g^2}{4\pi R}\bigg[\,
  \frac{q_{\rm UV}}{2}|\phi_{\rm UV}|^2 +\frac{q_{\rm IR}}{2}e^{-2k\pi R}
  |\phi_{\rm IR}|^2 +\xi_{\rm FI} (1-e^{-2k\pi R}) \bigg]^2 \nonumber \\
  && +\left|\frac{\partial W_{\rm UV}}{\partial \phi_{\rm UV}} \right|^2 
  +e^{-4k\pi R} \left|\frac{\partial W_{\rm IR}}{\partial \phi_{\rm IR}}
  \right|^2 +\frac{ke^{-4k\pi R}}{6M^3} \left|3W_{\rm IR}-\phi_{\rm IR}
  \frac{\partial W_{\rm IR}}{\partial \phi_{\rm IR}}\right|^2.
  \label{potential}
\end{eqnarray}
For later discussion, we present the equations of motion for the
auxiliary fields;
\begin{eqnarray}
  D \,&=& \frac{-g^2\xi_{\rm FI}}{2\pi R}(1-e^{-2k\pi R})
  -\frac{q_{{}_{\rm UV}}g^2}{4\pi R}|\phi_{\rm UV}|^2
  -\frac{q_{{}_{\rm IR}}g^2}{4\pi R}e^{-2k\pi R}|\phi_{\rm IR}|^2,\\[1mm]
  F_{\phi_{{}_{\rm UV}}}^\dagger \!\!&=& 
  -\frac{\partial W_{\rm UV}}{\partial \phi_{\rm UV}}, \\
  F_{\phi_{\rm IR}}^\dagger \!&=& -e^{-k\pi R}
  \frac{\partial W_{\rm IR}}{\partial \phi_{\rm IR}}
  +\frac{ke^{-k\pi R}}{6M^3}\, \phi_{\rm IR}^\dagger
  \Big(3W_{\rm IR}-\phi_{\rm IR}
  \frac{\partial W_{\rm IR}}{\partial \phi_{\rm IR}} \Big), \\
  F_T^\dagger \,&=& \frac{e^{-k\pi R}}{6\pi M^3}
  \Big(3W_{\rm IR}-\phi_{\rm IR}
  \frac{\partial W_{\rm IR}}{\partial \phi_{\rm IR}} \Big),
\end{eqnarray}
where we have simply assumed that the graviphoton field does not have
a nonzero expectation value. The scalar potential is a function of
boundary scalar fields and the modulus $R$. Assuming an appropriate
form of boundary superpotentials, we first minimize the potential
(\ref{potential}) with respect to matter scalars and find the minimum
value $V(R)$ of the potential, which generally depends on $R$. Then
doing the minimization of $V(R)$, we obtain the vacuum with a
stabilized value of $R$.

From the generic form of the scalar potential (\ref{potential}), we
have several observations for the radius stabilization: (i) First,
unless the FI term is present, the radius is not stabilized. For a 
vanishing value of $\xi_{\rm FI}$, there is a $D$-flat direction in
the potential. Along this direction, the potential has the $R$
dependences only in the form of $e^{-4k\pi R}$ and the vacuum goes to
infinity. (ii) Another observation is that, unless supersymmetry is
broken, the radius is not stabilized at a finite value. Generally
speaking, the scalar potential vanishes for unbroken supersymmetry and hence
cannot fix the radius. In the present case, the FI term itself gives
rise to a non-vanishing potential. However it is a monotonous function
of $R$ and consequently, the radius $R$ is fated to have a runaway
behavior.

We thus find that, for stabilizing the radius modulus, the FI term
must be present and also four-dimensional supersymmetry must be broken
(leading to non-vanishing vacuum energy). An interesting point is
that, in curved five-dimensional theory, a FI term is automatically
induced and can cause required supersymmetry breaking. Note that the
above arguments are applied to the models with boundary multiplets 
only, and therefore including hypermultiplets and/or bulk dynamics may
change the conclusion. For example, as in the model of Section 4,
non-trivial $R$-dependences of bulk-field wavefunctions 
generate $R$-dependent (super)potential terms, which can lead 
to (supersymmetric) radius stabilization. We have shown in the above
that this cannot be obtained by boundary matter only and supersymmetry
needs to be broken.

\subsection{Radius determination}

As we mentioned, four-dimensional supersymmetry must be broken to
stabilize the size of the compact fifth dimension. In the absence of
charged matter fields, the FI term leads to a non-vanishing potential
for the radius modulus, as discussed in Section 3. However it does not
have any minimum with respect to $R$ for anomaly-free theory. Moreover
it might be unfavorable with such a vacuum energy that supersymmetry
is broken at a high-energy scale. In this section, we are thus
interested in including the charged matter contribution. With a
non-vanishing FI term, the $D$-flatness condition points to the unique
vacuum with nonzero expectation values of charged matter
scalars. Therefore adding appropriate perturbation to superpotential
terms, the charged field directions are lifted and hence supersymmetry
is broken, as needed.

The most simple case is to introduce on the $y=0$ boundary a
vector-like chiral multiplets $\phi_{\rm UV}$ and $\bar\phi_{\rm UV}$
(with $U(1)$ charges $+q_{\rm UV}$ and $-q_{\rm UV}$ respectively) and
their mass term:\footnote{Other examples are to introduce a
vector-like chiral multiplets on the $y=\pi R$ boundary, to introduce
only a constant superpotential, and so on. We find that, in either of
these cases, the radius modulus is not stabilized at a finite value.}
\begin{equation}
  W_{\rm UV} \,=\, m\phi_{{}_{\rm UV}}\bar\phi_{{}_{\rm UV}}.
\end{equation}
Together with the $D$-term potential, one can see that supersymmetry
is broken by small perturbation ($m\sim$ TeV). It is however found
from (\ref{potential}) that the modulus potential in this case has the
maximum only. We thus incorporate a constant superpotential on 
the $y=\pi R$ boundary: $W_{\rm IR}=\omega$. Such a constant
superpotential can be obtained in various ways and here we do not
consider any details of its origin. Without loss of 
generality, $q_{{}_{\rm UV}}\xi_{\rm FI}$ is taken to be positive and
the vacuum is given by
\begin{equation}
  \phi_{\rm UV} =\, 0, \qquad |\bar\phi_{\rm UV}|^2 =\, 
  \frac{2\xi_{\rm FI}}{q_{\rm UV}} (1-e^{-2k\pi R}) 
  -\frac{8\pi Rm^2}{q_{\rm UV}^2g^2}.
  \label{vev}
\end{equation}
The vacuum energy, which depends on the radius $R$, becomes
\begin{equation}
  V(R) \,=\, \frac{2m^2}{q_{\rm UV}}\xi_{\rm FI}(1-e^{-2k\pi R})
  -\frac{4\pi Rm^4}{q_{\rm UV}^2g^2}+\frac{3k|\omega|^2}{2M^3}e^{-4k\pi R}.
  \label{VR}
\end{equation}
Minimizing the vacuum energy determines the value of $R$. We find the 
solutions to $\frac{\partial V(R)}{\partial R}=0$;
\begin{equation}
  e^{2k\pi R} \,=\, \frac{k\xi_{\rm FI} q_{{}_{\rm UV}}g^2}{2m^2}
  \bigg(1\pm\sqrt{1-\frac{6\omega^2}{g^2\xi_{\rm FI}^2M^3}}\;\bigg).
  \label{sol}
\end{equation}
In this vacuum with a stabilized $R$, supersymmetry is broken by 
nonzero $D$ and $F$ terms;
\begin{equation}
  D \,=\, \frac{-2m^2}{q_{\rm UV}}, \qquad F_{\phi_{\rm UV}}
  \,\simeq\, -m\sqrt{\xi_{\rm FI}/q_{\rm UV}}, \qquad 
  F_T \,=\, \frac{\omega^*e^{-k\pi R}}{2\pi M^3}.
  \label{DFF}
\end{equation}

\begin{figure}[htbp]
\begin{center}
\scalebox{1.3}{\includegraphics{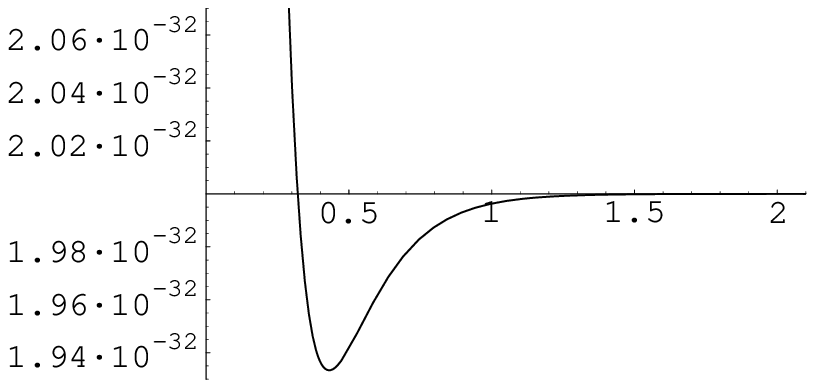}}
\put(8,81.5){$kR$}
\put(-237,167){$V(R)$}
\caption{The four-dimensional vacuum energy as a function of the size
of the fifth dimension (\ref{VR}). In the figure, we 
take $m=10^{-15}$, $\omega=10^{-16}$, $k=0.1$, and $\xi=0.01$ in the
unit of the fundamental scale $M$.}
\label{VRfig}
\end{center}
\end{figure}

There are two typical scales of the constant term $\omega$ such 
that $F_T/R$ is on the order of supersymmetry-breaking 
scale $O({\rm TeV})$. The first case is given by a suppressed value 
of $\omega$ compared to the fundamental scale $M^3$. In this case, it
is easily found by analyzing the second derivative of the potential
that the minimum of $V(R)$ is given by $e^{2k\pi R}\simeq 
\frac{3q_{\textrm{UV}}k|\omega|^2}{2m^2 \xi_{\textrm{FI}}M^3}$.
Therefore the radius is stabilized around the value
\begin{equation}
  kR \,\simeq\, O(1),
\end{equation}
and a large metric warp factor does not arise. The
supersymmetry-breaking contribution from the radius modulus is found
to be $F_T\,\sim\,\omega/M^3\ll 1$. Fig.~\ref{VRfig} shows an explicit
form of the vacuum energy (the potential of the radius modulus) for a
suppressed value of $\omega$.
The stabilized modulus obtains the mass squared $m_T^2=
\frac{e^{2k\pi R}}{6k\pi^2M^3}\frac{\partial^2 V(R)}{\partial R^2}$
which is always positive definite at the minimum. In the parameter
region we now consider, it is approximately given by
\begin{equation}
  m_T^2 \,\simeq\, \frac{4m^4\xi_{\rm FI}^2}{9q_{\rm UV}^2\omega^2}
  \;\simeq\; O\big({\rm (TeV)}^2\big),
\end{equation}
with the canonical kinetic term of the radius modulus.

\begin{figure}[htbp]
\begin{center}
\scalebox{1.2}{\includegraphics{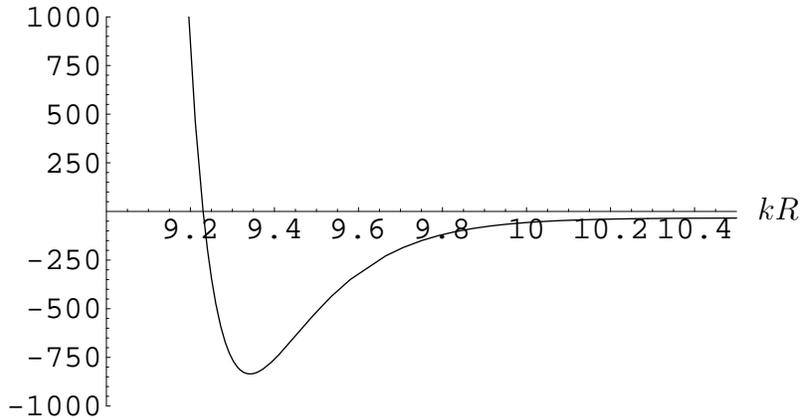}}
\put(8,84){$kR$}
\caption{Typical behavior of the vacuum 
energy $V(R)$ for $m=10^{-15}$, $\omega=0.01$, $k=0.1$, and $\xi=1$
in the unit of $M$. (The figure plots the potential from which we have
subtracted a radius-independent 
constant $\frac{m^2\xi_{\textrm{FI}}}{q_{\textrm{UV}}}$ and normalized
it by $\frac{4m^4}{kq_{\textrm{UV}}^2g^2}$.) \ For a smaller value 
of $\omega/\xi_{\rm FI}$, the valley of the minimum becomes steeper.} 
\label{VRfig2}
\end{center}
\end{figure}

Another typical scale of $\omega$ is a natural scale in the theory,
namely, $\omega\sim M^3$. In this case, the solution (\ref{sol}) means
that the minimum is around $e^{2k\pi R} \,\simeq\,
\frac{k\xi_{\textrm{FI}} q_{\textrm{UV}}g^2}{m^2}$ and therefore,
\begin{equation}
  kR \,\simeq\, \frac{1}{2\pi}
  \ln\Big(\frac{M}{\textrm{TeV}}\Big)^2 \;\,\sim\; 10.
\end{equation}
As a result, the metric warp factor gives significant effects, and
also we have a suppressed value of radius modulus $F$ 
term; $F_T\,\sim\,e^{-k\pi R}\,\sim\,\textrm{(TeV)}/M\,$. In
Fig.~\ref{VRfig2}, we show a typical behavior of the vacuum 
energy $V(R)$.
The mass of the radius modulus with the proper normalization in this
region is
\begin{equation}
  m_T^2 \,\simeq\, \frac{4m^4e^{2k\pi R}}{3q_{\rm UV}^2g^2M^3}
  \;\simeq\; O\big(\textrm{(TeV)}^2\big).
\end{equation}
Again we have a TeV-scale massive modulus field. For a large metric
factor, KK-excited modes have suppressed masses above $O(\textrm{TeV})$ 
in four-dimensional theory. It might require a careful treatment to
examine whether the effective Lagrangian of zero modes is valid in
this parameter region. In addition, the minimum might not be so steep
that it is not meta-stable within the cosmological evolution. However
a higher potential barrier can be achieved by a smaller value of the
constant superpotential, and a unstable vacuum is easily avoided.

\subsection{Towards realistic models}

We have shown that a nonzero FI term can stabilize the radius modulus to
a realistic value. As well known in four-dimensional theory, the FI
term is capable of explaining Yukawa hierarchy of quarks and leptons
and also of providing interesting sparticle spectrum. In this
subsection, we present a toy model towards constructing realistic
theory in higher dimensions where a single existence of FI term has
various important implications to phenomenology.

Let us consider one-generation `lepton' 
multiplets $\phi_L$, $\phi_R$, and `Higgs' $H$ as well as a
vector-like multiplets $\phi$ and $\bar\phi$. The latters play the 
radius stabilizer discussed in the previous section. For simplicity,
we focus only on the $U(1)$ factor and ignore the standard model gauge
groups. Incorporating these gauge factors is rather
straightforward. Now suppose that $\phi_L$ comes from a
five-dimensional hypermultiplet $(\phi_L,\bar\phi_L)$ with orbifold
parities $(+,-)$. So $\phi_L$ contains a massless mode in
four-dimensional effective theory. All other multiplets $\phi_R$, $H$,
$\phi$ and $\bar\phi$ are assumed to
be confined on the UV boundary. The $U(1)$ charges of these multiplets
are listed in the Table~1.
\begin{table}[htbp]
\centering
\begin{tabular}{c|ccccc}
& ($\phi_L,\bar\phi_L$) & $\phi_R$ & $H$ & $\phi$ & $\bar\phi$ \\ \hline
$U(1)$ & ($q_L,-q_L$) & $q_R$ & 0 & $1$ & $-1$
\end{tabular}
\caption{The $U(1)$ charge assignment. We take the charge of the Higgs
field zero, for simplicity, and the matter charges $q_L$ and $q_R$ are
positive.}
\end{table}
As in usual four-dimensional case, the $U(1)$ charges of matter fields
are taken as positive, which will be important to have non-vanishing
Yukawa couplings, positive sfermion masses squared, and also the
potential analysis in the previous section to be valid. The 
charge $q$ has been set to +1 without loss of any generalities. Note
that, with only these multiplets at hand, the zero-mode 
effective $U(1)$ theory is anomalous. But some anomaly cancellation
may easily be assumed, for example, introducing additional charged
multiplets. An important point here is that, even if effective
four-dimensional theory is anomaly free, an induced FI term can be
nonzero due to the curved extra dimension. Note also that there are no
gauge anomalies for the standard gauge groups, if included, provided
that anomalies are cancelled within massless
modes~\cite{ExDanomaly,FIexD2,FIexDwarp,ExDanomaly2}. Therefore in
case that light-mode spectrum is that of the standard model, we do not
worry about gauge (and gravitational) anomalies for any field
configurations in the extra dimension.

The FI term may be radiatively generated even for anomaly-free 
particle contents. In the $U(1)$ theory above, the one-loop
contribution is given by (\ref{FI}) 
with $\xi_{\rm UV}=\frac{(q_L+2q_R)\Lambda^2}{32\pi^2}$ 
and $\xi_{\rm IR}=\frac{-q_L\Lambda^2}{32\pi^2}$ where $\Lambda$ is
near the fundamental scale $M$. As seen in the previous section, the
successful radius stabilization needs a positive value 
of $\xi_{\rm IR}$, which is not satisfied in the present
form. ($\xi_{\rm UV}$ must also be positive to have Yukawa couplings
and supersymmetry breaking.) A simple way to cure this problem is to
introduce charged hypermultiplets which have even (odd) orbifold
parity at the UR (IR) boundary, or vice verse. These additional
multiplets contribute to the FI 
coefficients $\Delta\xi_{\rm UV}=\Delta\xi_{\rm IR}=
\frac{Q\Lambda^2}{32\pi^2}$ where $Q$ is the sum of $U(1)$ charges of
added multiplets. Moreover they contain no zero modes and do not
change the standard model spectrum. With this implementation, both
FI-term coefficients can safely be positive. In the previous 
analysis, $\xi_{\rm FI}$ is replaced with $\xi_{\rm IR}$ in the
solution (\ref{sol}) and the expressions of the radion mass, but in
the expectation values of $\bar\phi_{\rm UV}$ 
and $F_{\phi_{\rm UV}}$, $\xi_{\rm FI}$ is approximately given 
by $\xi_{\rm UV}$, which is suitably positive if $\xi_{\rm IR}$ is
made positive.

The Yukawa couplings for matter multiplets are described by
gauge-invariant higher-dimensional operators~\cite{FN} on 
the $y=0$ boundary:
\begin{equation}
  W_{\rm UV} \,=\, h\Big(\frac{\bar\phi}{M}\Big)^{q_L+q_R}\phi_L\phi_RH,
  \label{WY}
\end{equation}
where $h$ is the $O(1)$ coupling constant. If one included other
generations originated from bulk hypermultiplets, their Yukawa
couplings are allowed by bulk supersymmetry only on the boundaries
as (\ref{WY}). The potential analysis shows that only $\bar\phi$
develops a non-vanishing expectation value, i.e.\
(\ref{vev}).\footnote{The scalar $\bar\phi_L$ also has a 
negative $U(1)$ charge and might be worried to obtain a nonzero
expectation value. However the equation of motion for the $\phi_L$
scalar implies that this is not the case as long as there is no source
term of $\phi_L$ on the UV boundary.}
As a result, the above operator gives an effective Yukawa 
coupling $y\sim h\big(\frac{\langle\bar\phi\rangle}{M}\big)^{q_L+q_R}$. 
The expectation value of $|\bar\phi|^2$ is proportional to the FI term
and thus an one-loop order quantity. 
Therefore we obtain $\frac{\langle\bar\phi\rangle}{M}\equiv\lambda\sim O(0.1)$
that is just suitable for describing realistic Yukawa
hierarchies. Furthermore, in this model, there is an additional
possibility to have Yukawa suppression unlike in pure four-dimensional
theory. That is a wavefunction factor of bulk hypermultiplet zero
modes. The zero-mode wavefunction $\phi_{L_0}$ depends on its bulk
mass  $c$ and $U(1)$ charge, and is given by~\cite{FIexDwarp}
\begin{equation}
  \phi_{L_0} \,=\, 
  N_0\exp\Big[\Big(\frac{1}{2}-c\Big)k|y|+q_L ae^{2k|y|}\Big],
\end{equation}
with $N_0$ being the normalization constant determined 
by $\int\!dy|\phi_{L_0}|^2=1$ and roughly given 
by $N_0^2\simeq \frac{(1-2c)k}{e^{(1-2c)k\pi R}-1}$.
The present setup 
predicts $a=\frac{g^2\xi_{\textrm{IR}}}{8k}e^{-2k\pi R}$ derived from
the background expectation value of $\Sigma$, which is fixed by the
five-dimensional $D$-term equation. The second term in the bracket is
thus tiny in all region of the fifth dimension and can be dropped. We
then find that $N_0$ provides additional suppression of the effective
Yukawa coupling in the present model. For $c>\frac{1}{2}$, the
corresponding zero mode is localized at $y=0$ and yields no
suppression of Yukawa couplings which come from the operator on 
the $y=0$ boundary. On the other hand, the $c<\frac{1}{2}$ case gives
a Yukawa suppression by the 
factor $N_0\sim e^{(\frac{1}{2}-c)k\pi R}\ll 1$. This reflects the
fact that the zero mode is peaked at away from the $y=0$ boundary. In
the conformal limit $c=\frac{1}{2}$, the normalization constant $N_0$
becomes a volume-suppression factor $\frac{1}{\sqrt{\pi R}}$ as in the
case of flat extra dimension.

Supersymmetry-breaking spectrum is related to the radius stabilization
and Yukawa coupling structure. According to (\ref{DFF}), there are
three types of contributions to supersymmetry-breaking
parameters. Since we now introduced only boundary multiplets, the
equation of motion of $\Sigma$ in five dimensions implies that the
auxiliary field $D$ has a flat wavefunction in the fifth
direction. Therefore the $D$-term contribution is universal to all
charged scalars in the theory. The $F$ component of $\phi$ provides
soft masses and trilinear couplings of scalar fields from
superpotential and/or K\"ahler terms. It is found that they are
higher-dimensional operators and suppressed by powers 
of $\lambda=\frac{\langle\bar\phi\rangle}{M}$ compared to the 
leading $D$-term contribution. Ignoring these higher-dimensional
corrections,\footnote{Bulk fields with vanishing $U(1)$ charges, like
scalar top quark, might receive the dominant soft masses from
higher-dimensional K\"ahler terms involving $\phi$.} non-holomorphic
scalar masses are given by
\begin{eqnarray}
  m^2_L &=& -q_LD +\partial_T\partial_{\bar T}
  \ln |N_0(T,\bar T)|^2 |F_T|^2 \nonumber \\
  &\simeq& 2q_Lm^2 +\bigg|
  \frac{\big(c-\frac{1}{2}\big)k\pi R}{\sinh\big[\big(c-\frac{1}{2}\big)
    k\pi R\big]}\,\frac{F_T}{2R}\bigg|^2 \qquad\;\; 
  \textrm{(bulk scalars)}, \\
  m^2_R &=& 2q_Rm^2 \hspace*{4cm} 
  \textrm{($y=0$ boundary scalars)},
\end{eqnarray}
where $N_0(T,\bar T)$ is the appropriate superspace extension of the
normalization constant $N_0$. The first terms are the $D$-term
contributions which are positive definite ($q_{L,R}>0$). The second
term in the bulk scalar mass $m_L^2$ comes from the radius modulus $F$
term~(\ref{DFF}). We have dropped the $\Sigma$ contribution in $m^2_L$
since it is suppressed in the wavefunction factor $\phi_{L_0}$ as
discussed above. The boundary scalars at $y=0$ receive no $F_T$
contribution as seen from the Lagrangian (\ref{L4D}) and have rather
different spectrum than those of bulk scalar fields.

In case that quarks and leptons originate from bulk hypermultiplets,
the standard model gauge multiplets must also reside in the
five-dimensional bulk. Then the gauginos obtain supersymmetry-breaking
masses from two contributions; the radius modulus $F$ term and
higher-dimensional operators $\int\!d^2\theta\,c_i
\frac{\phi\bar\phi}{M^2}W^{\alpha i} W_\alpha^i$. The masses of
zero-mode gauginos $M_{1/2}^i$ are given by
\begin{equation}
  M_{1/2}^i \,=\, \frac{-c_ig^2\lambda^2m}{\pi R}+\frac{F_T}{2R}.
  \label{gauginomass}
\end{equation}

Let us concentrate on the vacuum with $kR\sim O(1)$. In this vacuum,
the radius modulus $F$ term is
\begin{equation}
  \frac{F_T}{R} \;\simeq\; \frac{\omega^*e^{-k\pi R}}{2\pi RM^3}
  \;\simeq\; \frac{\lambda m}{\sqrt{12\pi^2MR}} \;\ll\, m.
\end{equation}
For bulk scalar masses, the dominant part therefore comes from 
the $D$-term contribution and the spectrum is similar to
four-dimensional anomalous $U(1)$ models. The $D$-term contributed
scalar masses have rich phenomenological implications such as flavor
violation~\cite{LFV-D}. On the other hand, the two contributions to
gaugino masses in (\ref{gauginomass}) are comparable in size or 
the $F_T$ contribution can be dominant if low-energy effective theory
has a weak gauge coupling constant $\frac{g^2}{\pi R}\ll 1$. The
gauginos are found to have rather non-universal (non-unified) mass
spectrum in this scenario.

As seen in this toy model, the existence of FI term provides various
schemes to discuss phenomenological issues in higher-dimensional
theory. It can stabilize the sizes of extra dimensions, create Yukawa
hierarchies, and predict characteristic sparticle spectrum testified
in future particle experiments. Therefore more realistic model
construction along this line may deserve to be investigated.

\section{Summary}

In this work we have discussed the Fayet-Iliopoulos $D$ term as a
possible origin of radius stabilization in brane world models. We have
presented three different schemes for the stabilization. The simplest
case has no matter multiplets to stabilize the radius, but the theory
needs some cancellation mechanism of gravitational anomaly. The one of
the others contains bulk hypermultiplets, whose non-trivial
wavefunctions connect the two boundaries and then fix the size of the
extra dimension in terms of boundary couplings. On the other hand, the
third model involves only boundary dynamics and does not need the
presence of bulk matter fields. The radius is, in this case,
determined so that the vacuum energy is minimized after supersymmetry
breaking. Every scheme can lead to a significant warp factor or
near-flat metric, depending on the model parameters.

It should be noted that, in any model, the FI term is not a device
introduced just in order to stabilize the radius modulus. A
non-vanishing FI term is radiatively generated even if it is set to be
zero at classical level. An induced FI term depends on how charged
matter multiplets are distributed in the extra dimensions and is
therefore controllable. Moreover it is known in four-dimensional
models that the FI term is deeply connected with Yukawa hierarchy and
supersymmetry breaking. We have presented a toy model for fermion
Yukawa hierarchy correlated to the radius stabilization. The model
also predicts characteristic spectrum of sfermions and gauginos.

In the model we have drawn in Section 4, the bulk scalar fields
develop non-trivial wavefunction profiles in the extra dimension and
four-dimensional supersymmetry is unbroken. When included
supersymmetry breaking, it might give impacts on sparticle
spectroscopy and also deserve cosmological considerations. We leave
these phenomenological analysis to future investigations.

\bigskip

\subsection*{Acknowledgments}
The authors wish to thank H.~Abe, K.~Choi, T.~Hirayama and H.~Nakano  
for valuable discussions. T.~K.\/ is supported in part by the
Grant-in-Aid for Scientific Research  (\#16028211) and the
Grant-in-Aid for the 21st Century COE ``The Center for Diversity and
Universality in Physics'' from the Ministry of Education, Culture,
Sports, Science and Technology of Japan.

\end{document}